\documentclass[preprintnumbers, floatfix, preprintnumbers, letterpaper, twocolumn, superscriptaddress,nofootinbib]{revtex4}
\usepackage{graphicx}
\usepackage{microtype}
\usepackage{amsmath}
\usepackage{amssymb}
\usepackage{subfigure}
\usepackage{hyperref}
\usepackage{url}
\usepackage{xcolor}
\usepackage{color}
\usepackage{mathrsfs}
\usepackage{calrsfs}
\usepackage{amsfonts}
\usepackage{tabularx}
\usepackage{eucal}
\usepackage{latexsym}
\usepackage{ragged2e}
\usepackage{epsfig}
\usepackage{textcomp}
\usepackage{float}

\usepackage{caption}
\DeclareCaptionJustification{justified}{\leftskip=0pt \rightskip=0pt \parfillskip=0pt plus 1fil}
\definecolor{vividviolet}{rgb}{0.62, 0.0, 1.0}
\definecolor{amaranth}{rgb}{0.9, 0.17, 0.31}
\definecolor{palatinateblue}{rgb}{0.15, 0.23, 0.89}
\definecolor{brightpink}{rgb}{1.0, 0.0, 0.5}
\definecolor{cornflowerblue}{rgb}{0.39, 0.58, 0.93}
\definecolor{deepcarminepink}{rgb}{0.94, 0.19, 0.22}
\definecolor{radicalred}{rgb}{1.0, 0.21, 0.37}

\hypersetup{ linktoc=all,
	colorlinks, linkcolor={palatinateblue},
	citecolor={brightpink}, urlcolor={amaranth}
}

\graphicspath{{Images/}}

\def\sideremark#1{\ifvmode\leavevmode\fi\vadjust{\vbox to0pt{\vss% the remark
			\hbox to 0pt{\hskip\hsize\hskip1em%                          will appear only
				\vbox{\hsize1.3cm\tiny\raggedright\pretolerance10000%          on the side
					\noindent #1\hfill}\hss}\vbox to8pt{\vfil}\vss}}}%
\def\beq{\begin{equation}}
\def\eeq{\end{equation}}

\begin{document}
\title{How Barrow Entropy Modifies Gravity: With Comments on Tsallis Entropy}

\author{Sofia Di Gennaro}
\email{sofia.digennarox@gmail.com}
	\affiliation{Center for Gravitation and Cosmology, College of Physical Science and Technology, Yangzhou University, \\180 Siwangting Road, Yangzhou City, Jiangsu Province  225002, China}
	
\author{Hao \surname{Xu}}
\email{haoxu@yzu.edu.cn}
\affiliation{Center for Gravitation and Cosmology, College of Physical Science and Technology, Yangzhou University, \\180 Siwangting Road, Yangzhou City, Jiangsu Province  225002, China}

\author{Yen Chin \surname{Ong}}
\email{ycong@yzu.edu.cn}
\affiliation{Center for Gravitation and Cosmology, College of Physical Science and Technology, Yangzhou University, \\180 Siwangting Road, Yangzhou City, Jiangsu Province  225002, China}
\affiliation{Shanghai Frontier Science Center for Gravitational Wave Detection, School of Aeronautics and Astronautics, Shanghai Jiao Tong University, Shanghai 200240, China}

\begin{abstract}
Barrow proposed that the area law of the horizon entropy might receive a ``fractal correction'' $S\propto A^{1+\Delta/2}$ due to quantum gravitational effects, with $0\leqslant \Delta \leqslant 1$ measures the deviation from the standard area law. While such a modification has been widely studied in the literature, its corresponding theory of gravity has not been discussed. We follow Jacobson's approach to derive the modified gravity theory (interpreted as an effective theory), and find that in the stationary case the resulting theory only differs from general relativity by a re-scaled cosmological constant. Consequently in asymptotically flat stationary spacetimes the theory is identical to general relativity. The theory is not applicable when there is no horizon; the multi-horizon case is complicated. 
We emphasize on the importance of identifying the correct thermodynamic mass in a theory with modified thermodynamics to avoid inconsistencies. 
We also comment on the Hawking evaporation rate beyond the effective theory. In addition, we show that the Bekenstein bound is satisfied if the thermodynamic mass is used as the energy, up to a constant prefactor. We briefly comment on the Tsallis entropy case as well. Interestingly, for the latter, the requirement that Bekenstein bound holds imposes a lower bound on the non-extensive parameter: $\delta > 1/2$, which unfortunately rules out the previously suggested possibility that the expansion of the universe can accelerate with normal matter field alone.
\end{abstract} 

\maketitle

\section{Introduction: In Search of an Effective Gravity Theory from Barrow Entropy}

Barrow suggested that quantum gravity (QG) effects might introduce fractal structures on the horizon \cite{2004.09444}. Explicitly, he proposed that the Schwarzschild horizon is not a 2-sphere, but a ``sphereflake''. A sphereflake is obtained as follows: start with a Schwarzschild black hole whose Schwarzschild radius is $r_0$ and keep on adding more spherical surfaces to ``extend'' the horizon (similar to the construction of the well-known fractal Koch snowflake). At each iteration, add $N$ spheres of radius $r_{n+1}=\lambda r_n$, $\lambda<1$, to the existing spheres. Depending on the number of spheres added at each iteration and the scaling factor, the fractal dimension of the resulting sphereflake would be different. Since it is often much more convenient to have an explicit metric to work with, a spherically symmetric black hole metric in the usual coordinates $\left\{t,r,\theta,\phi\right\}$ that incorporates such an entropy correction was introduced and analyzed in the literature \cite{2110.07258,2205.07787}. However, this immediately raised a question: in what sense does such a metric describe the underlying geometry? After all, a sphereflake has a very intricate structure, is $r=\text{const.}$ supposed to correspond to a sphereflake? Does such a foliation even exist?

Motivated by the need to further understand such a metric, in this work, we try to construct an effective theory from the correction to the underlying thermodynamic. As we will argue, at the level of such an effective theory in which QG degrees of freedom are traced out, the fractal structure is not visible, so the usual foliation $\left\{t,r,\theta,\phi\right\}$ does make sense. However, it also turns out that in the vacuum case the theory is just general relativity (though perhaps with a modified gravitational constant).

Our approach will follow the landmark paper \cite{J} of Jacobson, in which he derived Einstein's field equations assuming the area law $S=\eta \mathcal{A}$ (Bekenstein-Hawking entropy has $\eta=1/4G$). The idea is to consider the heat flow across the horizon due to the energy carried by matter field (with energy-momentum tensor $T_{ab}$)
\begin{equation}
\delta Q =-\kappa \int_\mathcal{H} \lambda T_{ab} k^ak^b d\lambda d\mathcal{A},
\end{equation}
where $\mathcal{A}$ is the area of the horizon $\mathcal{H} $ with $k^a$ denoting the tangent vector of the horizon generators, $\kappa$ the surface gravity, and $\lambda$ is an appropriate affine parameter. 
The change of the horizon area is governed by the Raychaudhuri equation:
\begin{equation}\delta A = \int_\mathcal{H}  \theta d\lambda d\mathcal{A} =-\int_\mathcal{H}  \lambda R_{ab}k^ak^b d\lambda d\mathcal{A},
\end{equation} 
where $\theta$ is the expansion of the horizon generators. 
Then, using the first law of thermodynamics $\delta Q = T dS$, where $T=\kappa/2\pi$, we obtain
\begin{equation}\label{1}
-\kappa \int_\mathcal{H}  \lambda T_{ab} k^ak^b d\lambda d\mathcal{A}= -\frac{\kappa \eta}{2\pi} \int_\mathcal{H}  \lambda R_{ab}k^a k^b d\lambda d\mathcal{A}.
\end{equation}
This can only be valid if $T_{ab}k^ak^b=(\eta/2\pi)R_{ab}k^ak^b$ for all null vector $k^a$, thus $(2\pi/\eta)T_{ab}=R_{ab}+fg_{ab}$ for some function $f$. The contracted Bianchi identity then fixes $f$ and yields the Einstein field equations. Jacobson's work reveals a deep relationship between thermodynamics and gravity, thus opening a new field of research in gravitation. 

This also means that if the area law is modified, we should expect that the corresponding theory of gravity will no longer be general relativity (GR).
This brings us back to the Barrow entropy \cite{2004.09444}.
For a fixed mass, let $\mathcal{A}$ denote the actual horizon area. Without the Barrow modification, we would have a Schwarzschild black hole with area that we denote by $\mathcal{A}=A$.
Upon turning on the entropy correction, the original area $\mathcal{A}=A$ then receives a correction and becomes $\mathcal{A}=\alpha A^{1+\frac{\Delta}{2}}$, where $\Delta$ is the Barrow entropy index satisfying $0 \leqslant \Delta \leqslant 1$.
Here $\alpha$ is a constant that is usually taken to be order unity for simplicity. Thus we set its value to be $\alpha=1$ in the following. Strictly speaking, Barrow's proposal is \emph{not} a modification of the area law, since he still assumed that $S=\mathcal{A}/4$, only that $\mathcal{A}$ is now the area of a fractal surface (a sphereflake instead of $S^2$); note that because of the fractal nature the dimension of the area is no longer equal to 2 but larger. Therefore, naively, applying Jacobson's method one should expect that there is nothing new, that we should just obtain GR. This is true. However, as we will see, there can be a different interpretation.
 
\section{Barrow Gravity}

From Eq.(\ref{1}), the Barrow correction $\mathcal{A}=A^{1+\frac{\Delta}{2}}$ means we now have{\footnote{Note that for an actual fractal, we cannot define a tangent vector (at least not in the usual sense). For example, a Koch snowflake is evidently continuous but not differentiable. However, we are working at the level of effective theory in which the black hole surface is still smooth; see below.}}
\begin{equation}
\int_\mathcal{H} \lambda T_{ab} k^ak^b d\lambda dA^{1+\frac{\Delta}{2}} = \frac{\eta}{2\pi} \int_\mathcal{H} \lambda R_{ab}k^a k^b d\lambda dA^{1+\frac{\Delta}{2}},
\end{equation}
from which we can obtain\footnote{Mathematically this is just a change of variable from $\mathcal{A}$ to $A^{1+\frac{\Delta}{2}}$, but we emphasize that physically we are now treating -- as shall be argued below -- $A$ as the actual physical area (with the price that the entropy is no longer $A/4$). Thus, \emph{interpreted in this way}, Eq.(\ref{f5}) and Eq.(\ref{3}) are not simply re-labeling of Jacobson's equations but an extension of it.}
\begin{equation}\label{f5}
\int_\mathcal{H} \lambda \left(T_{ab}-\frac{\eta}{2\pi}R_{ab}\right)\left(1+\frac{\Delta}{2}\right)A^\frac{\Delta}{2}k^ak^b d\lambda dA = 0.
\end{equation}
Still assuming for now $\eta=1/4G$, we have\footnote{At this stage, we can also choose to recover GR by requiring just $T_{ab}-\frac{1}{8\pi G}R_{ab} = fg_{ab}$.} 
\begin{equation}\label{3}
\left(T_{ab}-\frac{1}{8\pi G}R_{ab}\right)\left(1+\frac{\Delta}{2}\right)A^{\frac{\Delta}{2}} = fg_{ab}
\end{equation}
for some function $f$. Taking the covariant derivative, and applying the contracted Bianchi identity \`a la Jacobson, we can eventually obtain
\begin{flalign}\label{full}
&\left(1+\frac{\Delta}{2}\right)A^{\frac{\Delta}{2}}\left[-\frac{1}{8\pi G}\nabla_b \left(\frac{R}{2}\right)\right] \notag \\ &+ \left(1+\frac{\Delta}{2}\right)\left(T_{ab}-\frac{1}{8\pi G}R_{ab}\right)\frac{\Delta}{2}A^{\frac{\Delta}{2}-1}\partial^a A = \partial_b f.
\end{flalign}
Here, we treat $\Delta$ as fixed, which is expected to be true over the horizon.
In the stationary case (horizon area is constant), we have only one term on the LHS: 
\begin{equation}\label{5}
\left(1+\frac{\Delta}{2}\right)A^{\frac{\Delta}{2}}\left[-\frac{1}{8\pi G}\nabla_b \left(\frac{R}{2}\right)\right] = \partial_b f,
\end{equation}
that is
\begin{equation}
f=\Lambda' -\frac{\left(1+\frac{\Delta}{2}\right)A^{\frac{\Delta}{2}}}{8\pi G} \frac{R}{2},
\end{equation}
for some constant $\Lambda'$.
Substituting this back into Eq.(\ref{3}) yields the Einstein equations for stationary spacetimes
\begin{equation}\label{EFE}
R_{ab} - \frac{1}{2}g_{ab}R + g_{ab} \frac{\Lambda}{\left(1+\frac{\Delta}{2}\right)A^{\frac{\Delta}{2}}} = 8\pi G T_{ab},
\end{equation}
where $\Lambda=8\pi G \Lambda'$ is the ``bare'' cosmological constant for $\Delta=0$. It may be more appropriate to consider the bare $\Lambda$ as a quantity whose physical dimensions vary with $\Delta$ so that overall the combination ${\Lambda}{\left[\left(1+\frac{\Delta}{2}\right)A^{\frac{\Delta}{2}}\right]^{-1}}$ has dimension of $\text{length}^{-2}$.
%%%%%

Thus, Barrow entropy yields a GR-like gravity but with a re-scaled cosmological constant\footnote{{We thank Manosh Manoharan for pointing out that Asghari and Sheykhi have previously already obtained Eq.(\ref{EFE}) in \cite{2110.00059}. However the focus of our work is quite different.}} $\tilde{\Lambda}:={\Lambda}{\left[\left(1+\frac{\Delta}{2}\right)A^{\frac{\Delta}{2}}\right]^{-1}}$. 
This seems to suggest that the theory can naturally accommodate a dynamical ``cosmological constant'', which would be advantageous in cosmology, though an analysis beyond the stationary assumption should be done to confirm this.
What is problematic is that the theory cannot be applicable to spacetimes with no horizon, such as a pure Anti-de Sitter spacetime since the derivation of the field equation requires a non-zero area. While this is true also for Jacobson's derivation of GR, it was not an issue since $A$ does not appear in the final field equation. Eq.(\ref{EFE}) is equally problematic for spacetimes that have more than one horizon -- how should the cosmological constant term be re-scaled if there are more than two areas? However, this is due to the fact that in the presence of multiple horizons, they each have an associated temperature, which makes ``heat flow across the horizon'' a delicate and complicated issue to begin with. The theory, at least in the form that it is derived in this work, is therefore also not applicable to multi-horizon spacetimes.

On the other hand, when the theory is applicable, if the bare $\Lambda$ is zero, then it is identical to GR.
Note that this is \emph{not} saying that Barrow is wrong about the fractalized Schwarzschild geometry, just that Jacobson's method only gives us an effective modified theory that does not exhibit all the QG properties. That is to say, the fractalized horizon is a QG effect that is not visible at the level of thermodynamics considered here. At the level of effective theory, we can therefore treat the horizon as a smooth surface still with area $A$, but the entropy may no longer equal a quarter of the horizon in general. This explains why we can have a modified gravity theory even though Barrow assumed $S=\mathcal{A}/4$ in the beginning.

It is worth noting that the Tsallis entropy \cite{Tsallis,Tsallis2,1903.03098}, when applied to black holes, is very similar to the Barrow entropy in form, being: $S_T=\left(\frac{A_0^{1-\delta}}{4G}\right) A^\delta$, where $\delta>0$ is the non-extensive parameter and $A_0$ a constant \cite{Tsallis2}; however, they are fundamentally different as Tsallis entropy replaces the underlying Boltzmann-Gibbs distribution with a non-extensive \emph{classical} generalization, whereas Barrow entropy has a QG origin. Using Tsallis entropy instead of the Barrow one, we found that Jacobson's method would lead to a modification of the Einstein equations on the matter side and acts as an effective gravitational constant $G_{\text{eff}}=\frac{G}{\delta} \left(\frac{A}{A_0}\right)^{1-\delta}$, which correctly reduces to the usual $G$ when $\delta=1$. Note that $G_{\text{eff}}\to \infty$ if $\delta \to 0$.

\section{Comment on the ``Modified'' Schwarzschild Solution}

Precisely because of the effective theory being oblivious to the fractal structures, we can consider a spherically symmetric solution. This was already implicitly assumed in the literature when the metric of the modified Schwarzschild black hole takes the standard static form \cite{2110.07258,2205.07787} $ds^2=-g(r)dt^2 + g(r)^{-1}dr^2 + r^2 d\Omega^2$. Otherwise, as briefly mentioned in the Introduction, one is faced with a peculiar question: is the exterior spacetime foliated by $S^2$ \emph{except} at the horizon which is a sphereflake (which is problematic since the horizon is not convex -- there will be points that are not contained in the spherical sections outside the horizon, for example), or does the foliation itself consist of a family of sphereflakes (in which case it is natural to consider $\Delta$ to be a function of the distance from the horizon, so that $\Delta \to 0$ asymptotically)? In either case the geometry $d\Omega^2$ is not trivial, and it is doubtful whether such a foliation even exists. From the effective theory point of view this complication does not arise -- $d\Omega^2$ is just the metric of a round sphere. It is likely that to properly describe the geometry of the fractalized black hole beyond the effective theory requires a vastly more complicated metric, if not full QG.
 
At the level of effective theory, however, since the vacuum theory is just GR, its spherically symmetric static solution must be Schwarzschild, so what is the implication of not having an area law?
Let the metric function be $g(r)=1-2GM/r$ as usual.
The Hawking radiation is 
\begin{equation}\label{hawking}
T=\frac{g'(r)}{4\pi} \bigg\rvert_{r_+} = \frac{1}{8\pi{GM}}.
\end{equation}

Note that this starting point already differs from the approach in  \cite{2110.07258,2205.07787}, in which the authors computed the Hawking temperature assuming the first law $dM=TdS$. The results were (with the units $G=1$)
\begin{flalign}
T=\frac{1}{(\Delta+2)(4\pi)^{1+\frac{\Delta}{2}}M^{1+\Delta}}, \\
g(r)=1-\frac{(\Delta + 2)M^{\Delta+1}(4\pi)^{\frac{\Delta}{2}}}{r} \label{wrong}.
\end{flalign}
This is problematic. First of all, it can be seen that the expression of $g(r)$ is just a re-definition of the Schwarzschild metric with ADM mass $\mathcal{M}$ equals to $[(\Delta + 2)/2](4\pi)^{\frac{\Delta}{2}}{M}^{\Delta+1}$, yet in the first law the mass is $M$. In principle this is not necessarily a problem, since it is possible that the thermodynamic mass differs from the ADM mass \cite{1411.0833}. To see where problems might arise, we should think about this from a more operational point of view. Consider actually observing a black hole and measuring its mass, this would give us, essentially\footnote{With the usual caveat that the ADM mass is technically defined at spatial infinity.}, $\mathcal{M}$. From here we have no knowledge of what $M$ and $\Delta$ are separately, they are not observable quantities. In other words, the metric in Eq.(\ref{wrong})  \emph{cannot} be interpreted as a modified Schwarzschild solution; \emph{it is Schwarzschild}. Thus, one cannot constrain  $\Delta$ by observations if we use a vacuum asymptotically flat metric. Furthermore, the black hole entropy should be prescribed only after we know what the measured mass is, because in order to compute the entropy one must have the horizon radius, which is given by the metric. This means that we cannot even begin the derivation from the first law, as we need the metric in the first place to provide $r_h$ that goes into the entropy $S \propto (r_h^2)^{1+ {\Delta}/2}$. In the derivation of \cite{2110.07258}, $r_h$ was taken from the usual Schwarzschild metric $r_h=2M$, not the actual metric.

From the effective theory point of view, as we have argued, the geometry is precisely Schwarzschild, with temperature given by the usual Eq.(\ref{hawking}). The astrophysically measured ADM mass is $M$, 
and the horizon radius expression that goes into Barrow entropy is $r_h=2M$, so there is no ambiguity. The only compromise one must make is that in such a theory the thermodynamic mass cannot be the same as $M$. 
It is helpful at this stage to restore all the factors of $\hbar, G, c$ and $k_B$ and consider the first law $dE=TdS$. Denote the thermodynamic mass as $M_{\text{therm}}$. Then
\begin{flalign}
&\int \frac{1}{T} d(M_\text{therm} c^2)= \int dS \\ &= \left(\frac{4\pi G^2M^2}{c^4}\right)^{1+\frac{\Delta}{2}}\left(\frac{k_B c^3}{G\hbar}\right)^{1+\frac{\Delta}{2}} \cdot C(\Delta),
\end{flalign}
where 
\begin{equation}
C(\Delta):=\left(\frac{k_B G}{c\hbar}\right)^{-\frac{\Delta}{2}}\cdot M_\text{Pl}^{-\Delta}
\end{equation}
is a dimensionful constant that needs to be included to keep the correct physical dimension for entropy. We have factored out the power of Planck mass $M_\text{Pl}^{-\Delta}$ for convenience. This is the only term that survives after collecting and canceling all the powers of $\hbar, G, c, k_B$.
We can then obtain
\begin{equation}
\int M ~dM_{\text{therm}} = 2^{\Delta-1} \pi^{\frac{\Delta}{2}}M^{2+\Delta} M_\text{Pl}^{-\Delta}.
\end{equation}
By the fundamental theorem of calculus, and assuming that $\Delta$ is fixed, we get
\begin{equation}
M = \frac{d}{dM_{\text{therm}}}(M^{2+\Delta}) 2^{\Delta-1}\pi^{\frac{\Delta}{2}}M_\text{Pl}^{-\Delta}.
\end{equation}
Separating variables and integrating finally yields
\begin{equation}\label{Mtherm}
M_{\text{therm}} = \frac{2^{\Delta-1}\pi^{\frac{\Delta}{2}}(\Delta+2)}{\Delta+1}M^{\Delta+1} M_\text{Pl}^{-\Delta}.
\end{equation}
This makes it clear that both $M_{\text{therm}}$ and $M$ have the same physical dimension of a mass; they are the same only when $\Delta=0$.

\section{Comment on the Bekenstein Bound}

Shortly after our work appeared on the arXiv, Abreu and Neto studied the Bekenstein bound assuming non-standard entropies including that of Barrow \cite{2207.13652}.
They showed that if one uses $M$ as the energy $E$, then the Bekenstein bound is violated since, with the units $M_\text{Pl}=1$, the inequality $S \leqslant 2\pi R E$ leads to $S \leqslant S^{\frac{2}{2+\Delta}}$, which is not valid for large black holes with $S>1$. In fact, with $R=2M$, the putative bound is equivalent to $E \geqslant 2^\Delta \pi^{\Delta/2}M^{1+\Delta}$. The expression on the RHS is exactly of the form of the thermodynamic mass in Eq.(\ref{Mtherm}), up to a factor of $(1/2)(\Delta + 2)/(\Delta+1)$, which lies in the range 0.75 and 1. Thus, if we identify $E=M_{\text{therm}}$, while the Bekenstein bound is still violated, the violation is ``only'' by this factor. Thus, we can save the Bekenstein bound by postulating a minor modification: $S \leq C \pi R E$, where $C=4(\Delta+1)/(\Delta+2)$ is a constant that contains $\Delta$. This then reduces to the usual Bekenstein bound when $\Delta$ goes to zero. The crucial physics in Bekenstein bound is contained in the entropy $S$ being bounded above by $R\times E$ for any fixed $\Delta$, the exact coefficient does not really matter. Thus we interpret this result as a non-violation of the Bekenstein bound.
This example further supports our case that one should be careful with identifying the correct thermodynamic mass.\\

As an additional evidence for this, let us repeat the calculation for the Tsallis entropy:
\begin{equation}
	S_T = \left(\frac{A_0^{1-\delta}}{4G}\right) A^\delta.
\end{equation}
The energy, which is the thermodynamic mass, is
\begin{equation}
	E=M_{\text{therm}} = \left(\frac{A_0}{\pi G^2}\right)^{1-\delta} 2^{4\delta-4} \frac{\delta}{2\delta-1}M^{2\delta-1}
\end{equation} 
The resulting equation is:
\begin{equation}
	S_T \leqslant A_0^{1-\delta} 2^{4\delta-2} G^{2\delta-1} \frac{\delta}{2\delta-1} \pi^\delta M^{2\delta}.
\end{equation}
We see that, contrary to what is argued in \cite{2207.13652}, the Bekenstein bound $S \leqslant 2\pi R E$ is satisfied if we take into account the thermodynamic mass, provided that $1/2 < \delta \leqslant 1$. 
The upper bound is tighter than the bound $\delta \leqslant 2$ obtained by considering the modified Friedmann cosmology in \cite{1806.03996}.
If we allow a constant pre-factor so that $S \leqslant C\pi RE$ as per the case of Barrow entropy discussed above, however, we can relax the bound to $\delta > 1/2$. 

Interestingly, we note that $\delta < 1/2$ in cosmology allows the universe to expand with only normal matter (i.e., with an equation of state $w\geqslant 0$) \cite{1806.03996}. Our analysis suggests that this is not feasible if we choose to keep a Bekenstein bound. Note that $\delta < 1/2$ also corresponds to $E=M_{\text{therm}} < 0$. Thus, one may be tempted to conclude that it is the positivity of the mass that imposes $\delta < 1/2$. For black holes, with $T$ and $S$ both positive, $E<0$ indeed does not make physical sense, and so the condition that $E<0$ is equivalent to that of having a Bekenstein bound.
However, thermodynamic energy (``internal energy'') can in principle be negative in a generic thermodynamical system in which the work done by the system is greater than heat given to the system.

\section{Comment on the Hawking Rate}

Because the black hole is identically Schwarzschild, the Hawking rate is unmodified \emph{at the level of effective theory}. Beyond the effective theory regime, when the fractal structure becomes important, it is not clear if we can still model Hawking evaporation by the simple Stefan-Boltzmann law (or even a somewhat modified version). Even if we could, there are many subtleties that have no clear answer. 
First of all, we remark that since Hawking particles are created outside the horizon, the emission rate is only modified if we accept that $\Delta$-correction also applies to other spacetime regions, not just on the horizon (this does not mean that other surfaces obey thermodynamical laws \cite{2207.04390}). This is reasonable: $\Delta$ might depend on the strength of the gravitational field and goes to zero asymptotically (see also the Discussion section for the running $\Delta$ case). 

Perhaps the most important question is: what is the power in the temperature in the Stefan-Boltzmann law when the horizon is fractalized? Statistical mechanics and field theory on fractals have been studied in the literature but many open questions remain \cite{1010.1148, 1210.6763, wang}.
Also, since a fractalized horizon is no longer a convex surface, not all particles emitted close to the horizon is headed ``outward'', this would also affect the Hawking rate. 
Note also that in \cite{2004.09444}, Barrow wrote that the Hawking rate goes up as we add more and more spheres in the construction of the sphereflake horizon, and in fact the rate could diverge if the final area diverges (see also \cite{neto}). However, it is not obvious that this is the right ``area'' to consider. Take a space-filling curve (such as the Hilbert curve) for example, whose image is the unit square. At each iteration of the construction of the fractal, more and more line segments were added, so that the length at the $N$-th step is $2^N-2^{-N}$. The final \emph{length} clearly diverges. But the final fractal is a unit square with (Hausdorff) dimension 2, whose area is 1. Likewise, if the final sphereflake horizon has dimension $r^{2+\Delta}$, and the entropy scales like the highest dimensional area of the emerging fractal, the emission surface should not be the sum of the 2-dimensional areas. Perhaps the approaches based on fractional quantum mechanics can be useful to understand this issue \cite{2107.04789}.

\section{Discussion}

Following Jacobson's derivation of GR from Bekenstein-Hawking area law, we have derived the effective modified gravity theory that corresponds to Barrow entropy. In the stationary case the theory is just GR but with a re-scaled cosmological constant. At the level of effective theory, the fractal structures are not visible. Instead, the theory treats the black hole as possessing a smooth spherical horizon, but satisfying a non-area law entropy in general. If we assume a static and vacuum spacetime, the black hole solution must be none other than Schwarzschild, but with a different thermodynamic mass. In general we must be very careful with identifying the correct thermodynamic mass that goes into the first law of black hole thermodynamics (see also \cite{1506.01248}). In most scenarios we expect the thermodynamics mass to be the same as the ADM mass, however from the literature we know this is not necessarily the case \cite{1411.0833}. If we insist on physical ground following \cite{2109.05315} that the thermodynamic mass should be the same as the ADM one (essentially by considering the conservation of energy of the collapse of a mass shell), then we have to conclude that the effective Barrow gravity is not viable. 
Even if we allow $M_{\text{therm}} \neq M$ (unlikely, but not impossible, since energy conservation is a delicate issue in a fully dynamical spacetime with no time translational symmetry), the overall message of our work, as well as 
\cite{2109.05315}, is that we should be careful when dealing with modified black hole thermodynamics in order to avoid inconsistencies (see also \cite{2207.07905}).

Let us now comment on the field equation Eq.(\ref{EFE}). Though it is tempting to think that the effective cosmological constant $\tilde{\Lambda}:={\Lambda}{\left[\left(1+\frac{\Delta}{2}\right)A^{\frac{\Delta}{2}}\right]^{-1}}$ might ameliorate the cosmological constant problem by allowing $\tilde{\Lambda}$ to be extremely small even if the bare $\Lambda$ is of its natural scale, one should keep in mind that Eq.(\ref{EFE}) assumes stationarity so that the horizon does not evolve. For cosmological spacetimes, the apparent horizon typically is a function of cosmic time, and thus one must consider the full Eq.(\ref{full}), which is highly nontrivial to analyze. Another interesting point we can notice is that since a Kerr-AdS black hole has a smaller area than a Schwarzschild-AdS black hole of the same physical mass, they also correspond to two different effective cosmological constants. In other words, the effective AdS length scale depends on the angular momentum. Furthermore, since the Kerr-AdS metric contains the AdS scale in various coefficients, this would also mean that the metric depends on the angular momentum in a vastly more complicated manner than it already does in GR. On the other hand, the fact that the effective cosmological constant naturally changes with the area (and thus with the black hole parameters) is in line with -- or better accommodates -- the idea of extended black hole thermodynamics \cite{1209.1272,0904.2765}, or ``black hole chemistry'' \cite{1404.2126,1608.06147}. To study the variation beyond a stationary horizon, is however, like in the cosmological setting, a complicated task.

Another possibility is that $\eta$ in $S=\eta \mathcal{A}$ may no longer be $1/4G$ but instead $\Delta$-dependent. This will modify the RHS of the field equation (i.e., give rise to an effective gravitational constant). In \cite{2205.09311}, it is argued why the Barrow index $\Delta$ is expected to run, i.e., energy-scale dependent, but since $\Delta$ is fixed over a given horizon, this does not affect the derivation of the field equation itself. The effective theory is incomplete in the sense that if $\Delta$ does run, the dynamics of $\Delta$ cannot be determined. This should be contrast with, e.g., Brans-Dicke theory in which the scalar field $\phi$ that acts as an effective gravitational constant appears not only in the modified Einstein's field equations but is also determined by its own equation of motion $(\Box \phi \propto T^a_{~a})$. This is not surprising as $\Delta$ is a QG correction -- its dynamics requires the detailed knowledge of the underlying QG theory, beyond the effective thermodynamic treatments. 

Finally, we emphasize that there are still a lot of subtle issues worth investigating when thermodynamics is modified. 
For example, it has recently been argued that when the Barrow entropy or the Tsallis entropy is used, one should replace the Hawking temperature with an effective ``equilibrium temperature''. One could ask how this may affect our analysis. However, given the ongoing debate \cite{1774277,1814105,1866306,2210.00324} we feel that any further investigation along this line would deserve a separate work.

\begin{acknowledgments}
HX thanks the Natural Science Foundation of the Jiangsu Higher Education Institutions of China (No.20KJD140001) for funding support. YCO thanks the National Natural Science Foundation of China (No.11922508) for funding support. YCO also thanks Brett McInnes for discussions.
\end{acknowledgments}

\end{document}